\documentclass[preprint,12pt,onecolumn]{article}\oddsidemargin=-0.54cm
\usepackage{natbib}
\usepackage{setspace}
\usepackage{xcolor}
\usepackage{amsmath}
\usepackage{graphicx}
\usepackage{subfig}
\usepackage{float}
\usepackage{times}
\usepackage{multicol}
\usepackage{hyperref}
\usepackage{xcolor}
\usepackage{authblk}

\hypersetup{
    colorlinks,
    citecolor=blue,
    filecolor=blue,
    linkcolor=cyan,
    urlcolor=blue
}

\begin{document}
\pagestyle{empty}
\onecolumn
\begin{center}

{\bf \Large Astrophysical constraints on the simulation hypothesis for this Universe: why it is (nearly) impossible that we live in a simulation}\\*[3mm]
\end{center}
\begin{center}
Franco Vazza (\footnotesize 1,2\normalsize)\\
\bigskip

\footnotesize

{\it  1 ~Dipartimento ~di ~Fisica ~e ~Astronomia, Università ~di~Bologna, ~Via ~Gobetti ~93/2, ~40122, ~Bologna, ~Italy;
 
\it 2 ~INAF, ~Istituto~di ~Radioastronomia ~di~ Bologna, ~via ~Gobetti ~101, ~41029 ~Bologna, ~Italy}
\normalsize
\end{center}

\section*{Abstract}
% write abstract here
We assess how physically realistic the "simulation hypothesis" for this Universe is, based on physical constraints arising from the link between information and energy, and on known astrophysical constraints. 
We investigate three cases: the simulation of the entire visible Universe, the simulation of Earth only, or a low resolution simulation of Earth, compatible with high-energy neutrino observations. 
In all cases, the amounts of energy or power required by any version of the simulation hypothesis are entirely incompatible with physics, or  (literally) astronomically large, even in the lowest resolution case. Only universes with very different physical properties can produce some version of this Universe as a simulation. On the other hand, our results show that it is just impossible that this Universe is simulated by a universe sharing the same properties, regardless of technological advancements of the far future. 
%\newpage
%\twocolumn
%\end{multicols}

\section*{Keywords}cosmology, information, simulations, black holes 
%\end{abstract}

\section{Introduction}

The "simulation hypothesis" (SH) is a radical and thought-provoking idea, with ancient and noble philosophical roots (for example, in works by  \citealt{Descartes1984-DESMOF} and \citealt{Berkeley1734-BERATC-6}) and frequent echos in the modern literature, which postulates that the reality we perceive is the creation of a computer program.

The modern version of the debate is usually referred to an influential article by \citet{Bostrom2003}, although several coeval science fiction movies contributed to make this theme extremely popular {\footnote{For example, it was the central topic of the 2016 Isaac Asimov Memorial Debate, featuring several astrophysicists and physicists: \url{https://www.youtube.com/watch?v=wgSZA3NPpBs}. This theme still periodically surges and generate a lot of "noise" in the web, both indicating the fascination it rightfully produce on the public, as well as the difficulty to debunk it.}}.

Despite its immense popularity, this topic has been rarely investigated in a scientific way, also because at first sight it might seem to be entirely out of the boundaries of falsifiability, and hence relegated to social media buzz and noise. 

A remarkable exception is the work by 
\citet{2014EPJA...50..148B}, who investigated the potentially observable consequences of the SH, by exploring the particular case of a cubic space-time lattice. They found that the most stringent bound on the inverse lattice spacing of the universe is $\sim 10^{-11} \rm ~GeV^{-1}$, derived from the high-energy cut off of the cosmic ray spectrum.  Interestingly, they proposed that the SH can be tested through the  distributions of arrival direction of the  the highest energy cosmic rays, through the detection of a degree of rotational symmetry breaking, associated with the the structure of the underlying lattice. Our work will also use ultra high energy cosmic rays and neutrinos to put constraints on the SH, even if in a totally different way (e.g. Sec.\ref{low_res_earth}).

Our starting point is that "Information is physical" \citep[][]{1999PhyA..263...63L,1996PhLA..217..188L}, and hence any numerical computation requires a certain amount of power, energy, computing time, and the laws of physics can clearly tell us what is possible to simulate, and under which conditions. 
We can use these simple concepts to assess the physical plausibility- or impossibility - of a simulation reproducing the Universe \footnote{As a sign of respect for what this astrophysicist honestly thinks is the only "real" Universe, which is the one we observe through telescopes, I will still use the capital letter for this Universe - even if for the sake of the argument this is at times assumed to be just a simulation, being run in another universe.}   we live within, and even of some lower resolution version of it. 
%This particular astrophysicists is interested to explore whether, using  basic laws of physics and information theory, it is in principle possible to  derive strong constraints for the simulation hypothesis (SH for short in the reminder of the paper). 

%constrain the amount of computation which is needed to advance a realistic simulation of the Universe we live in.
Moreover, the powerful physical nature of information processing allows us to even sketch the properties than any other universe simulating us must have, in order for the simulation to be feasible. 

This paper is organised as follows: Section~\ref{methods} presents the quantitative framework which we will use to guess the information and energy budget of simulations; Sec.\ref{results} will present our results for different cases of the SH, and Sec.\ref{discussion} will critically assess some of the open issues in our treatment. Our conclusions are finally summarised in Sec.\ref{conclusions}.

%Therefore, even in the provocative and extreme scenario in which the Universe we live in is entirely the result of an advanced simulation, we can use the information content of the simulation to constrain some properties of the other universe in which a (super)computer is running the simulation. As we shall see, the amount of energies and power required by all possible versions of the SH are (literally) astronomical, and in this sense we shall use astrophysical knowledge to assess how much likely it is that such simulations are performed in another universe. 

%which is supposedly running our reality on a computer.
%The result of such exercise, which is manifestly subject to many orders of magnitude uncertainties, invariably shows that under no circumstances (unless the physical laws of the universe simulating us are entirely different from the known ones) the collective human experimentation of the Universe can be the result of a sophisticated simulation. 

\section{Methods: the holographic principle and information-energy equivalence}
\label{methods}

In order to assess which resources are needed to simulate a given system, we need to quantify how much information can be encoded in a given portion of the Universe (or in its totality). 
The Holographic Principle (HP) arguably represents the most powerful tool to establish this connection. It was inspired by the modelling of black hole thermodynamics through the Bekenstein bound (see below), according to which the maximum entropy of a system scales within its encompassing surface, rather than its enclosing volume. The HP is at the core of the powerful "AdS/CFT" correspondence (where AdS stands for anti-de Sitter spaces and CFT for conformal field theories)  which links string theory with gravity in five dimensions, with the quantum field theory of particles with no gravity, in a four-dimensional space \citep[e.g.][]{1998AdTMP...2..231M}. 
%was proposed between the anti-de Sitter spaces (AdS) used in quantum gravity, and conformal field theories (CFT) that are quantum field theories inspired to the  Yang–Mills theories to describe elementary particles.
%While the experimental proofs (or its disputation) of the HP are still debated, its close connection with the theory of black holes makes (whose testing .....
According to the HP, a stable and asymptotically flat spacetime region with boundary of area $A$ is fully described by no more
than $A/4$ degrees of freedom, or about 1 bit of information per Planck area, defined as:

\begin{equation}
l_p^2 = G \hbar/c^3 = 2.59 \cdot 10^{-66} \rm cm^2. 
\label{lp}
\end{equation}

where obviously $l_p=\sqrt{\hbar G/c^3} \sim 1.6 \cdot 10^{-33} \rm cm$ is the 
Planck scale.  
While in a local classical field theory description
there are far more degrees of freedom, the  excitation of more than $A/4$ of these degrees of freedom will trigger gravitational collapse \citep[e.g.][for reviews]{2002RvMP...74..825B,2004ConPh..45...31B,2005bhis.book.....S}.
The total entropy contained with the holographic area $A$ follows from the  generalised second law of thermodynamics, giving the Bekenstein bound, which applies to systems which are not strongly self-gravitating and whose energy is $E \approx Mc^2$: 

\begin{equation}
S \leq 2 \pi k_B \frac{E ~R}{\hbar ~c}=A/4
\end{equation}

where $R$ is  the circumferential radius of the smallest sphere that fits around the matter system, assuming (nearly) Euclidean spacetime for simplicity, and $M$ is the mass of the system. 

%(assuming that gravity is sufficiently weak to allow for a
%choice of time slicing such that the matter system is at
%rest and space is almost Euclidean).
%Applied to the full size of the Universe,  this yields the astounding maximum information capacity of $\sim 10^{100} \rm bits$ which can be stored in the entire fabric of space-time using all available Planck lengths thereby contained   \citep[e.g.][]{2003SciAm.289b..58B}.  

Next, we can use the classical information-entropy equivalence \citep[][]{1929ZPhy...53..840S}, which states that the minimum entropy production connected with any 1-bit measurement is given by:

\begin{equation}
    H  = k_B  \rm log (2)
\end{equation}

where the $\rm log(2)$ reflects the binary decision.  The amount of entropy for a single 1-bit measurement (or for any single 1-bit flip, i.e. a computing operation) can be greater than this fundamental amount, but not smaller, otherwise  the second law of thermodynamics would be violated.

Therefore, the total information (in [bits]) that can possibly be encoded within the holographic surface $A$ is:

\begin{equation}
I_{\rm max} = \frac{S}{H} = 2 \pi \frac{E ~R}{\hbar ~c \log(2)}. 
\label{Imax}
\end{equation}

How much energy is required to encode an arbitrary amount of information ($I$) in {\it any} physical device? Computation and thermodynamics are closely related: in any irreversible \footnote{It must be noted that also {\it reversible} computation, with no delation of bits or dissipation of energy,  is possible.  However, irreversible computation is unavoidable both for several many-to-one logical operations (AND or ERASE) as well as for error correction, in which several erroneous states are mapped into a single correct state \citep[e.g.][]{Sandberg1999-SANTPO-19,2000Natur.406.1047L}.} computation, information must be erased, which is a physical reduction of the number of possible states of one physical register to $0$. This necessarily leads to the decrease of entropy of the device used for computing, which {\it must} be balanced with an equal (or larger) increase in entropy of the Universe. Thermodynamics thus leads to the necessity of dissipating energy (via heat) to erase bits of information.
The related cost of erasing one single bit is thus given by the "Brioullin's inequality": %, e.g. ][]{2000Natur.406.1047L}:

\begin{equation}
\Delta E \geq k_B T \log(2)
\label{DeltaE}
\end{equation}

where $T$ is the (absolute) temperature where the energy gets dissipated.  If we apply this to the maximum information within the holographic sphere, we get:

\begin{equation}
E_{\rm I} = 2 \pi k_B T \frac{E R}{\hbar c }.
\end{equation}
It is interesting to compute the ratio between the enclosed energy within the holographic surface, $E$, and the energy required to encode the same information, $E_I$, which scales as 

\begin{equation}
 \chi = \frac{E_I}{E} \approx 27.5 \frac{T}{1^\circ  \rm K} \frac{E}{1 \rm erg}  \approx 0.04 \frac{T}{1^\circ \rm K} \frac{M}{m_p} 
\end{equation}
This physically means  even at the low temperature of $\sim 1^\circ K$ for computing,  the minimum energy required to fully describe all internal degrees of freedom of system becomes larger than the actual energy that the system contains, already for systems with a mass $M \sim 25 m_p$ ,  or more.
This is enlightening, as it shows how the full simulation of any macroscopic (or astronomical) object is bound to require an astoundingly large amount of energy, which in turn allows us to assess the plausibility of the SH  against known astrophysical bounds.

The exact bounds derived from Eq.\ref{Imax} depend on 
what systems are considered and on how $R$, $E$, and $S$ are defined \citep[e.g.][]{Page1982947,2002RvMP...74..825B}. It must also be stressed that a  key assumption in Bekenstein's derivation formula is that the gravitational self–interaction of the system can be neglected, as it highlighted by the fact that the Newton constant $G$ does not appear. Although in the application explored in this work this assumption is reasonably verified, we notice that the estimates for black holes can significantly differ \citep[e.g.][and references therein]{2002RvMP...74..825B}.

\section{Results: Information requirements and energy bounds}
\label{results}

\subsection{Full simulation of the visible Universe}
\label{full_res_universe}
Based on the above formulas, we can first estimate the total information contained within the holografic surface with radius equal to the observable radius of the Universe, $R_U \approx 14.3 \rm Gpc$ (comoving), and assuming a critical density $\rho_c=8.5 \cdot 10^{-30} \rm g/cm^3$, which is valid for a flat Universe at $z=0$ and using a value for the Hubble constant $H_0=67.4 \rm ~km/s/Mpc$. 
This gives a total energy

\begin{equation}
E_U = \frac{4 \pi}{3} R^3_{U} \rho_c c^2 \sim 2.7 \cdot 10^{78} \rm erg
\end{equation}

Based on Eq.\ref{Imax} this results into a maximum information of 

\begin{equation}
I_{\rm U} \sim 3.5 \cdot 10^{124} \rm bits
\end{equation}

which requires
\begin{equation}
E_{\rm I, U} \sim 8.9 \cdot 10^{108} \rm erg
\end{equation}

of encoding energy, assuming a computing temperature equal to the Microwave Background temperature nowadays ($T_{\rm CMB} =2.7^\circ (1+z) \rm K $).

As anticipated, since $E_{\rm I,U} \gg E_U$, there simply is not enough energy within the entire observable Universe to simulate another similar universe down to the Planck scale, in the sense that there are not even remotely available resources to store the data and even begin the simulation. 
Therefore, the SH applied to the entire Universe is entirely  rejected, based on the energy beyond imagination it requires. 

As noted above, slightly different estimates for the  total information content of the Universe can been obtained replacing the Bekenstein bound with the Hawking-Bekenstein formula, which computes the entropy of the Universe if that would be converted into a black hole:

\begin{equation}
I'_{\rm max} = \frac{GM^2}{\hbar c} \sim 2.0 \cdot 
10^{124} \rm bits
\label{eq:ImaxBH}
\end{equation}
which is in line with similar estimates in the recent literature (\citealt[e.g.][once we rescaled the previous formula for the volume within the cosmic event horizon, rather than from the volume within the observable Universe]{2010ApJ...710.1825E}; see also \citealt{2024arXiv241211282P} for a more recent estimate), and obviously incredibly larger than the amount of information which is generated even by the most challenging "cosmological" simulations produced to-date in astrophysics (e.g. $\sim 1.6 \cdot 10^{13} ~\rm bits$ of raw data in the Illustris-1 simulation, \citealt[][]{2014Natur.509..177V}).

{\bf This leads to the FIRST CONCLUSION: simulating the entirety of our visible Universe at full resolution (i.e. down to the Planck scale) is physically impossible.}

\bigskip

\subsection{Full simulation of planet Earth}
Next, we  can apply the same logic to compute the total memory information needed to describe  our planet ($R_{\oplus}=6.37 \cdot 10^{8} \rm cm$ , $M_{\oplus}=5.9 \cdot 10^{27} \rm g$ and $E_{\oplus}=M_{\oplus} c^2=5.46 \cdot 10^{48} \rm erg$). We assume  here that a planet is the smallest system that "the simulator" must model to recreate the daily experience that humankind collectively considers the reality \footnote{For the sake of the argument, since we surmise here the existence of a  skilled simulator which somehow can use computing centres as large a planet-sized black hole,  we can concede with no difficulty that it can also create a consistent simulation of the experience of $\leq 10^3$ astronauts that have temporarily left our planet and have probed a larger portion of space. The simulator must also use a  modest amount of extra computing resources to produce fake data to constantly keep astrophysicists and cosmologists busy while they think they study the Universe.}.

Again based on Eq.\ref{Imax} we get

\begin{equation}
I_{\rm max,\oplus} = 9.81 \cdot 10^{74} \rm bits
\label{imax_earth}
\end{equation}

and for the total energy needed to code this information:

\begin{equation}
E_{\rm max, \oplus} = 2.55 \cdot 10^{59} \rm erg
\end{equation}

assuming, very optimistically (as it will be discussed later) that $T=T_{\rm CMB}$. 
The amount of energy required {\it to even start} the simulation of our planet is enormous, and it can be easily put into astrophysical context: 
\begin{itemize}

\item this is of the same order of the rest mass energy of globular clusters, like Palomar2  ($M_{gc} \sim 3.3 \cdot 10^5 M_{\odot}$, e.g. \citealt{2018MNRAS.478.1520B}): $E_{\rm rm, gc}=M_{\rm gc}c^2 \approx 5.6 \cdot 10^{59} \rm erg$;
\item this is also of the order of the potential binding energy of the halo of our Galaxy ($M_{\rm gal} \sim 1.3 \cdot 10^{12} M_\odot$ and $R_{\rm gal}=287 \rm kpc$, e.g. \citealt{2019A&A...621A..56P}): $U_{\rm rm, gal} \sim 3GM_{\rm gal}^2/(5R_{\rm gal}) \sim 3.1 \cdot 10^{59} \rm ~erg$.

\end{itemize}

Therefore, the initialisation of a complete simulation of "just" a planet like Earth requires either to convert into energy the entire stellar mass of a typical globular cluster, or to use the equivalent energy necessary to unbind all stars and matter components in the Milky Way. 

Elaborating on the  implausibility of such a simulation, based on its energy cost, is straightforward: 
while this is indeed the requirement just to begin the simulation, roughly the same amount of energy needs to be dissipated for each timestep of the simulation. This means  that already after $\sim 10^6$ timesteps, the required energy is equivalent to the entire rest mass energy of the Milky Way, or roughly to the total potential energy of the most massive clusters of galaxies in the Universe. 

Moreover, while the minimum mass required to contain this information corresponds to a black hole mass prescribed by Eq.\ref{eq:ImaxBH} , which gives $\sim 0.32 M_{\odot}$ using $I_{\rm max,\oplus}$ , the actual energy required to encode the entire amount of bits,  $E_{\rm max,\oplus}$,  can at most be confined at within a radius corresponding to the Schwarzschild radius given by:

\begin{equation}
R_S = \frac{2~GE_{\rm max,\oplus}}{c^4}  \sim 4.95 \cdot 10^{9} \rm cm 
\end{equation}

which is 70\% of the radius of Jupiter ($R_J=6.99 \cdot 10^{9} \rm cm$), meaning that planetary-sized computer must be deployed for this simulation. The equivalent mass enclosed within such radius will be of course very large: $M_{\rm max,\oplus}=E_{\rm max,\oplus}/c^2=1.68 \cdot 10^5 M_{\odot}$. 
Moreover, such computing Jupiter must be continuously supplied with a similar amount of energy for time step, while all dissipated energy is somehow released outside of the system (and even without raising the computing temperature), which makes this scenario even more implausible.

Interestingly, such planetary-sized computers were theoretically explored by \citet{Sandberg1999-SANTPO-19}, who presented a through study of all practical limitations connected to heat dissipation, computing power, connectivity and bandwidth, arriving to a typical estimate of $\sim 10^{47} \rm bits$ for a realistic Jupiter-sized computer. This is impressive, and yet $27$ orders of magnitude fewer than what is required to encode the maximum information within the holographic surface containing our planet. Even setting aside the tremendous distance between hypothetical and realistic memory capacity of such colossal computer, the  next problem is that concentrating so much mass and energy in such a limited volume will inevitably produce levels of high energy emission and heating, as in standard black holes and their related accretion discs. 

If a black hole actively accretes matter, the 
kinetic temperature acquired by accreted matter is very large: 

\begin{equation}
T_{\rm acc} \sim  \frac{G M_{\rm max,\oplus} m_p}{3 R_J ~k_B}\sim \frac {2 m_p ~c^2}{3~ k_B},
\label{Tacc}
\end{equation}

i.e. regardless of the actual mass and radius of the black hole into consideration, the temperature acquired by accreted particles is in the $\sim (m/m_p)  \cdot 10^{7} \rm K$ regime, for a generic particle with $m/m_p$ relative to the proton mass.  Such  temperature is manifestly much larger than the very  optimistic CMB temperature we previously assumed to compute the {\it minimum} energy necessary to encode the full information of the simulation, i.e. $T_{\rm acc} \sim 10^7 T_{\rm CMB}$.  From $E \propto k_B T I_{\rm MAX}$ (Eq.\ref{Imax}) it follows that the actual energy requirement is a factor $\sim 10^7$ larger, even in the most optimistic configuration, requiring now a computer with radius $\sim 5 \cdot 10^{17} \rm cm$, i.e. $\sim 0.16$ parsecs. 

{\bf This leads to a  SECOND CONCLUSION:  
simulating planet Earth at the full resolution (i.e.  down to the Planck scale) is practically impossible as it requires to access to a galactic amount of energy.}

\begin{table*}
    \centering
    \begin{tabular}{|c|c|c|c|c|} \hline 
         SIMULATION&  SYSTEM SIZE& MIN. RESOLUTION& MIN. INFORMATION&   MIN. ENERGY\\
 & [cm] & [cm] & [bits]& [erg]\\\hline \hline 
         Full Res. Universe& $4.47 \cdot 10^{31}$ & $1.64\cdot 10^{-33}$ & $3.5 \cdot 10^{124}$&  $8.9 \cdot 10^{108}$ \\ \hline 
         Full Res. Earth& $6.37 \cdot 10^{8}$ &$1.64\cdot 10^{-33}$ & $9.8 \cdot 10^{74}$&  $3.0 \cdot 10^{59}$ \\ \hline 
 Low Res. Earth&$6.37 \cdot 10^{8}$ & $1.24 \cdot 10^{-21}$ &$1.65 \cdot 10^{51}$& $4.3 \cdot 10^{35}$\\ \hline
    \end{tabular}
    \caption{Summary table of the tested simulation hypothesis, and of their memory and energy requirements.}
    \label{tab}
\end{table*}

\bigskip

\subsection{Low-resolution simulations of planet Earth}
\label{low_res_earth}

Next, we shall explore the possibility of "partial" or "low resolution" simulations of planet Earth, in which the simulation must only resolve scales which are routinely probed by human experiments or observations, while using some sort of "subgrid" physics for any smaller scale. 

The Planck scale $l_p$ is at the core of physics as we know it, but it is not a directly measurable scale by any means. In the framework of the SH, it is well conceivable that $l_p$ appears in the equations of our physics, but that the simulation instead effectively works by discretising our reality at a much coarser scale, thus providing a "low-resolution" simulation \footnote{As a matter of fact, in almost any conceivable numerical simulation which is run in physics, the Planck constant and its associated length scale $l_p$ are at the core of many physical relations, and yet the effective spatial resolution of simulations is coarser than by many orders of magnitude.}.

%The problem here is that we have to imagine how the holographic bound should apply, for a reality in which the smallest scale is not the Planck lenght, but a larger one. 

What is the smallest scale, $\Delta x_{\rm min}$ that a realistic simulation of our Planet should resolve, in order for the simulation not to be incompatible with available human experiments or observations done on the Planet? 

High energy physics, through the De Broglie relation, $\lambda=hc/E \sim h/p$, establishes an effective  link between observed energy particle phenomena and spatial length-scales. 
So the smallest resolution that the simulation must have depends on the highest energies which are routinely probed by human experiments. 
On Earth, the smallest high energy scale probed by man made experiments is in $\sim 10^{-16} \rm  cm$ ballpark based on the particle collision energy scale reached with the Large Hadron Collider \citep[e.g.][for a recent review]{2025arXiv250104714S}. However, this is true only limited to the portion of the terrestrial surface where the LHC is deployed (i.e. a ring of 27 km at most 200 m below the sea level).  

A much  larger energy scale is probed by the  detection of Ultra High Energy Cosmic Rays (UHECRs) as they cross our atmosphere and trigger the formation of observable Cherenkov radiation and fluorescent light \citep[e.g.][]{2014JCAP...10..020A,2015APh....64...49A}. Observed UHECRs are characterized by a power-law distribution of events, and the largest energy ever recorded for an UHECR is $\sim 3 \cdot 10^{20} \rm eV$  (1991, "Oh my god particle", e.g. \citealt{1995ApJ...441..144B}) while the second probably exceeded the $\sim 2.4 \cdot 10^{20} \rm eV$ range (2022, "Amaterasu" particle, e.g. \citealt{Unger_2024}). The precise determination of very energetic is subject to absolute energy calibration uncertainties, also due to their uncertain composition, hence we can use $E_{\rm UHECR}=10^{20} \rm ~eV=1.6 \cdot 10^{8} \rm erg$ as a conservative limit:  this yields a length scale $\lambda_{\rm UHECR} \sim 1.2 \cdot 10^{-24} \rm cm$. However, a hardcore proponent of the SH might still argue that this spatial scale is not really  the most stringent experimental limit on the minimum spatial scale of the global Earth simulation, because UHECRs probe only the last $\sim 10^2 \rm km$ of air above the ground level, where particle cascades and airshowers are triggered by the interaction between UHECRs and air molecules.  Hence, the majority of the simulated volume of our Planet might still  be kept a much lower level of spatial resolution by the simulator, in order to limit computing resources. 
And indeed, the best resolution which is currently sampled for the interior of Earth, based on the advanced analysis of seismic shear waves diffracting along the core-mantle boundary, is "just" of order $\sim 10^2 \rm km$ \citep[e.g.][]{JENKINS2021116885,GEO}. 

Luckily, high energy astrophysics still comes to the rescue, due to the detection of very high energy extragalactic neutrinos {\it crossing the inside of our planet}.
Since the first discovery of energetic neutrinos by the IceCube observatory \citep[][]{2013Sci...342E...1I,2013PhRvL.111b1103A}, the existence of a background radiation of neutrinos, likely of extragalactic origin, in the  $10  \rm TeV- 2 PeV$ energy range, has been firmly established. 
Such neutrinos are very likely produced in external galaxies, even if the exact mechanism is highly uncertain \citep[e.g.][for recent works]{buson23,neronov24,padovani24}. 
In any case, the maximum energy scale probed so far reaches about  $E_\nu \sim 10^{17} \rm eV$ \citep[][]{2025Natur.638..376K}, i.e. lower than the maximum energy reached by UHECRs but in line with theoretical expectations about the likely hadronic production mechanisms at the source.
Unlike UHECRs, however, neutrinos cross our planet entirely, since they are extremely weakly interacting with anything else. As a matter of fact, the IceCube observatory is more sensitive to events which are produced {\it on the other side of the planet}, with respect to its location on the South Pole as this reduces the contamination by lower energy neutrinos of atmospheric or solar origin.   Therefore, the established detection of this event can be used to constrain the minimum length scale which must be effectively adopted by any low-resolution simulation of our planet: $\lambda_\nu=hc/E_{\nu} \sim 1.2 \cdot 10^{-21} \rm cm$ \footnote{Especially because we focus here on the simulation of Earth, a reasonable question would be whether such low resolution simulation would still be capable of reproducing the observed biological process on the Planet. A $\sim 10^{-21} \rm cm$ length seems to be safely smaller than any process known to be relevant for biology, and thus a hypothetical simulation using an efficient sub-grid modelling of processes below such resolution might correctly reproduce biology on Earth. If instead some biological processes will be shown to depend on $< 10^{-21} \rm cm$ scales, this can be used to revise our constraints on the smallest scale to be resolved by the simulation, and call for an even more implausibly large amount of energy or power.}. A simulation with an effective resolution coarser than this will be incompatible with our experimental data on neutrinos.  
Less straightforward is how to use this knowledge to estimate the information required for such reduced resolution simulation of planet Earth, using the holographic principle as before.  The most conservative choice, appears to be to still apply the HP approach, but rescaling the application of Eq.\ref{Imax} to the  minimum element of area possible in the low resolution simulation  ($\propto \lambda_\nu^2$), instead of the Planck area (Eq.\ref{lp}).  Therefore, the minimum necessary information to be encoded  follows from rescaling Eq.\ref{imax_earth} for the ratio of the two areas: 

\begin{equation}
I_{\rm \oplus,low} \approx I_{\rm max, \oplus} \cdot \frac {l_p^2}{\lambda_\nu^2} \approx 1.65 \cdot 10^{51} \rm bits.
\label{eq:Imaxlow}
\end{equation}

This gives a minimum encoding energy

\begin{equation}
E_{\rm \oplus,low} = 4.31 \cdot 10^{35} \rm erg
\end{equation}

if we very conservatively use  $T=T_{\rm CMB}$ as above (which we are going to relax later on). 

%However, there is no obvious reason why the Bekenstein bound should still apply here, considering that the fundamental scale $l_p$ is replaced with another more arbitrary scale. On the other hand, a more classical estimate based on a volumetric view of information also causes problems:

%\begin{equation}
%I_{\rm \oplus,classical} \approx \frac{4 \pi R_{\oplus}^3}{3 \lambda_\nu^3}  \approx 5.67 \cdot 10^{89} \rm bits
%\label{eq:Imaxlow}
%\end{equation}

%i.e. a much larger information content than what we derived in Eq. \ref{imax_earth}, which is the maximum allowed with $R_{\oplus}$ based on the holographic principle indeed. 

At the face value, this energy requirement is far less astronomical than the previous one: it corresponds to the conversion into energy of $\approx 2.4 \cdot 10^{-19} M_{\odot}$, or $\approx 7.9 \cdot 10^{-14} M_{\oplus}$ ($\sim 4.8 \cdot 10^{14} \rm g$). This is still equal to the total energy radiated by the Sun in two minutes, considering the solar constant ($L_{\odot}\sim 3.8 \cdot 10^{33} \rm erg/s$), yet it is an amount of energy which a fairly advanced civilisation might possibly access to.  

Eq.\ref{eq:ImaxBH} gives the size of the minimum black hole capable of storing $I_{\rm \oplus,low}$: $M_{\rm BH,low}=4.4 \cdot 10^{-13} M_{\odot}$. The radius relative to this mass is also fairly small: $R_{\rm BH,low}=1.3 \cdot 10^{-7} \rm cm$.  Therefore, while the total energy required for the initial encoding of the simulation's data is  still immense by modern human standards, it is tiny in astrophysical terms.  However, next we are going to show that the only way to process the data of such simulation to advance it at a large enough speed, is the access to unattainably large computing power, which  makes this last low resolution scenario impossible too. 

We largely refer here to the seminal work by \citet{2000Natur.406.1047L} for the computing capabilities of black holes. 
The ultimate compression limit of a black hole can provide in principle the most performing computing configuration for any simulation. Thus by showing that not even in this case the simulation can be performed, we can argue about its physical impossibility. Based on the classical picture of black holes, no information is allowed to escape from within the event horizon. However, the quantum mechanical view is different as it allows, through the emission of  Hawking radiation as black holes evaporate, to transfer some information on the outside. 
Following \citet{2000Natur.406.1047L}, even black holes may theoretically be programmed to encode the information to be processed within their horizon. Then an external observer, by examining the correlations in the Hawking radiation emitted as they evaporate, might be able to retrieve the 
the result of the simulation outside. 
Even such a tiny black hole has a very long evaporation timescale: $t_{\rm ev} \sim G^2 M_{\rm BH,low}^3/(3 \hbar c^4 \rm k)$, with the constant $\rm k \sim 10^{-3}-10^{-2}$ depending on the species of particles composing the bulk of the BH mass, which gives $t_{\rm ev} \geq 10^{35} \rm s$. 

The temperature relative to the Hawking radiation for such a black hole is given by: 

\begin{equation}
T_H \sim \frac{\hbar c^3}{8 \pi G M_{\rm BH, low} k_B}  \sim 1.4 \cdot 10^5 \rm K.
\end{equation}
%The next stringent problem is posed by the processing of this information, to actually run the simulation. To see what the problem is, a small detour into the physics of information processing is in order. 
%The actual employed power by the supercomputer or facility used to run the simulation depends on the computing rate used by the simulator.  We can apply the laws  of physics s to constrain the configuration and energy requirements of the simulation reproducing our reality in the low-resolution Earth case.  

Quantum mechanics constrains the maximum rate at which a system  can move from one distinguishable quantum state to another. A quantum state with average energy $\bar{E}$ needs a time of order (at least) $\delta t \sim  \pi \hbar/2\bar{E}$ to evolve into another orthogonal and distinguishable state, and hence  the  number of logical operations per unit time (i.e. the computing frequency) that can be performed by a computing device is the inverse of the timescale from the Heisenberg uncertainty principle: $f \sim 1/\delta t \sim 2\bar{E} /(\pi \hbar)$. 
If such black hole calculator uses all its storage memory, its maximum computing power {\it per bit} (i.e. the number of logical operations for each single bit) is estimated to be: 

\begin{equation}
N_{\rm op} = \frac{k_B 2 \log(2) \bar{E} }{\pi \hbar S}   
\label{Nop}
\end{equation}

where $S$ is the black hole entropy.   The link between thermodynamic entropy and temperature here is given by  $T= (\partial S/\partial \bar{E})^{-1}$. By integrating the relationship linking $T$, $S$ and $\bar{E}$ we get $T =C \bar{E}/S$, in which $C$ is a constant of order unity, which depends on the actual medium being used (e.g. $C=3/2$ for an ideal gas or $C=4/3$ for photons of a black-body spectrum). The dependence between the computing power and the working temperature is manifest: "the entropy governs the amount of information the system can register and the temperature governs the number of operations per bit per second it can perform", as beautifully put by  \citet{2000Natur.406.1047L}. 

Estimating the working temperature of such a device is not obvious.
As an upper bound, we can use the temperature of accreted material at the event horizon of an astrophysical black hole from  Eq.\ref{Tacc} ($T \sim 10^7 \rm K$) and get:

\begin{equation}
N_{\rm op} \approx  \frac{k_B 2 \log(2) T }{\pi \hbar}  \sim 5.6 \cdot 10^{17}[\rm operations/bit/s].
\label{Nop1}
\end{equation}

By multiplying for the total number of bits encoded in such a black hole, we get its total maximum computing  power:

\begin{equation}
    P_{\rm op} \sim N_{\rm op} \cdot I_{\oplus,low} \sim  9.5 \cdot 10^{68} [\rm bits/s],
    \label{pop1}
\end{equation}

On the other hand,  if the black hole does not accrete matter, the lowest temperature at the even horizon is the temperature relative to the Hawking radiation: $T_H \sim 1.4 \cdot 10^5 \rm K$ for this mass. We thus get: 

\begin{equation}
N'_{\rm op} \approx 8.0 \cdot 10^{15}[\rm operations/bit/s].
\label{Nop2}
\end{equation}

In this case, we get for the total computing power:

\begin{equation}
    P'_{\rm op} \sim N'_{\rm op} \cdot I_{\oplus,low} \sim  1.3 \cdot 10^{67} [\rm bits/s].  
    \label{pop2}
\end{equation}

This computing power may seem immense, yet it is not enough to advance the low resolution simulation of planet Earth in a reasonable wall-clock time.
We notice that the minimum timestep that the simulation must resolve in order to consistently propagate the highest energy neutrinos we observe on Earth is 
 $\Delta t \approx \lambda_\nu/c \sim 4.1 \cdot 10^{-32} \rm s$. In the extremely conservative hypothesis that only a few operations per bit are necessary to advance every bit of the simulation forward in time for a $\Delta t$ timestep, $\sim O(10^{31})$ operations on every bit as necessary for the simulation to cover just $1$ second of evolution of our Universe. The two previous cases instead can at best achieve $\sim    8 \cdot 10^{15}-5.6 \cdot 10^{17}$ operations per bit per second, depending on the working temperature. 
In that conditions,  a single second in the low resolution simulation of planet Earth requires: 

\begin{itemize}
\item $t_{\rm CPU } \sim 4.2 \cdot 10^{13} \rm s$ of computing time, i.e. $\sim 1.4 \cdot 10^{7} \rm yr$, using the computing power given by Eq.\ref{Nop1}; 
\item $t_{\rm CPU} \sim 3.0 \cdot 10^{15} \rm s$ of computing time, i.e.  $\sim 1 \cdot 10^8 \rm yr$ using the computing power given by  Eq.\ref{Nop2},
\end{itemize}

which are in both cases absurdly long wall clock times. 
Therefore,  an additional speed-up of order $\times 10^{15}-10^{17}$ (or larger) would be necessary in to advance the low resolution simulation of Earth faster than real time \footnote{It should also be considered that the time dilatation effect of General Relativity will also introduce an additional delay factor between what is computed by the black hole, and what is received outside, although the amount of the delay depends on where exactly the computation happens.}.  In this case, the necessary computing power will be just impossible to  collect: 

\begin{itemize}
\item $dE/dt \sim 1.1 \cdot 10^{73} \rm erg/s$ if the working temperature is $\sim 10^5 \rm K$,
\item $dE/dt \sim 9.4 \cdot 10^{74} \rm erg/s$ if the working temperature is $\sim 10^7 \rm K$,
\end{itemize}

which means to convert into energy many more than all stars in all galaxies within the visible Universe (and using a black hole with mass $\sim 10^{-13} M_{\odot}$ for this). No known process can even remotely approach this power, and thus also this scenario appears absurd. 
 
{\bf This leads us to the THIRD CONCLUSION: even the lowest possible resolution simulation of Earth (at a scale compatible with experimental data) requires geologically-long timescales, making it entirely implausible for any purpose.}

%And this, under the extremely ideal conditions of a) perfect parallel computing; b) negligibly small error rate, despite the high noise at these temperature (see e.g. \citealt{2000Natur.406.1047L} for a discussion). 

\section{Discussion}
\label{discussion}

Needless to say, in such a murky physical investigation several assumptions can be questioned, and a few alternative models can be explored.
Here we review a few ones which appears to be relevant, even if it is anticipated that the enthusiasts of the SH will probably find other ways out.  
\bigskip

\subsection{Can highly parallel computing make the simulation of the low resolution Earth possible?}

A reasonable question would be whether performing highly parallel computing could significantly reduce the computing time estimated at the end of Sec.\ref{low_res_earth}. In general, if the computation to perform is serial, the energy can be concentrated in particular parts of the computer, while if it is 
 parallelisable, the energy can be spread out evenly among the different parts of the computer. 
 
The communication time across the black hole horizon ($t_{\rm com} \sim  2R/c$) is of the same order of the time to flip a single bit ($\sim \pi \hbar/(2 \bar{E})$, see above), hence in principle highly parallel computing can be used here.  However, the somewhat counter-intuitive result of \citet{2000Natur.406.1047L} is that the energy $E$ is divided among $N_{\rm proc}$ processing units (each operating at a rate $\sim 2E/(\pi \hbar N_{\rm proc})$) the total number of operations per second performed by the black hole remains the same: $\sim N_{\rm proc} 2E/(\pi \hbar N_{\rm proc}) =2E/(\pi \hbar)$, as in Sec.\ref{low_res_earth}. This strictly follows from the quantum relation between computing time and spread in energy explored in Sec.\ref{low_res_earth}. Thus if the energy is allocated to more parallel processor, the energy spread on which they operate gets smaller, and hence they run in a proportionally slower way.  Finally,  if the computing is spread to a configuration significantly less dense than a black hole,  by keeping the same mass, higher levels of parallelisation can be used, but the computing time will increase as a black hole computer already provides the highest number of operations per bits per second (Eq.\ref{Nop}).

%here is that for computing at the even horizon scale of a black hole (which still is the most efficient computing device we can possibly think of, given the enormous concentration it allows) the computation must be fully serial as every bit coded on the black hole will be essentially connected to every other bit over the course of any single computing step. 
%This has been 

\bigskip

\subsection{What if the time stepping is not the one given by neutrino observations?}
What if (for reasons beyond what our physics can explain) high energy neutrinos can be accurately propagated in the simulation with a time stepping much coarser than the one prescribed by $\Delta t=\lambda_\nu/c \sim 4.1 \cdot 10^{-32} \rm s$? In this case, the next constraint to fulfil is still very small, and  given by the smallest time interval that humans could directly measure so far in the laboratory: this is $\Delta t' \approx 10^{-20} \rm s$ \citep[e.g.][]{10.1038/nphys3941}. In this case, applying the same logic of Sec.\ref{low_res_earth} we get $t_{\rm CPU} \sim 40-10^4 \rm s$ depending on the computing temperature, which still means a simulation running several orders of magnitude slower than the real time. 
Moreover, even in this scenario the amount of power to process is (literally) out of this world: $dE/dt \sim k_B T \log(2) \cdot P_{\rm op} \sim 10^{44}-10^{47} \rm erg/s$ if we rescale Eq.\ref{pop1}-\ref{pop2} for the new time rate.  This power is astronomically large but not entirely impossible: 
mergers between massive clusters of galaxies, which are among the most energetic exchanges of matter in the Universe, "only" produce $\sim 10^{45} \rm erg/s$ of power \citep[e.g.][]{mv07}. 
Distant quasars can radiate energy up to $\sim 10^{47} \rm erg/s$ \citep[e.g.][]{2024ApJ...960..126V}, while 
supernovae can release up to $\sim 10^{52} \rm erg/s$, mostly in the form of energetic neutrinos and only in the the first $\sim 1-10$ seconds of their explosion. Gamma Ray Burst from Hypernovae can dissipate up to $\sim 10^{54} \rm erg/s$ on timescales of days \citep[e.g.][]{2014MNRAS.443...67M}. Finally the detection of gravitational waves by merging black holes (with masses of several tens of $M_{\odot}$) probed energy dissipation rates of several  $\sim 10^{56} \rm erg/s$, but limited to the short timescale of the coalescence \citep[e.g.][]{PhysRevLett.116.241102}. In anycase, conveying in a steady way such gigantic amount of energy through the microscopic black hole required for this very low resolution scenario appears as an impossible task. 

%In essence, even an extremely low-resolution simulation of our Planet will require to use amounts of energy which in this Universe are dissipated either only the most extreme and disruptive astrophysical events that we know (in which entire stars or black holes are destroyed), or are dissipated on scales incredibly larger than the "small" black hole device we considered in Sec.\ref{low_res_earth}, which still is the most efficient we can consider here. 
\bigskip

\subsection{What about quantum computing?}

By exploiting the fundamental quantum properties of superposition and entanglement, quantum computers are more performing than classical computers for a large variety of mathematical operations \citep[e.g.][]{365701}. 
In principle, quantum algorithms are more efficient in using memory and they may reduce time and energy requirements \citep[e.g.][for a recent take]{chen2023quantum}, by implementing 
exponentially fewer steps compared to classical programs. 
However, these important advantages compared to classical computers 
do not change the problems connected with the SH analysed in this work, which solely arise from the relation between spacetime, information density and energy. According to the HP, the information bound used here is the maximum allowed within a give holographic surface with radius $R$, {\it regardless} of the actual technique used to reach this concentration of information, or to process it.  Moreover, our estimated computing power does not stem from the extrapolation of the current technological performances of classical computing, but it already represents  the maximum possible performance, obtained in the futuristic scenario in which a black hole can be used as ultimate computing device. 
In summary, while quantum computing might in principle be the actual way to get to the maximum computing speed physically allowed at the scale of black holes (or in any other less extreme computing devices), this technology will not be able to beat the currently understood limits posed by physics. 

\bigskip

\subsection{What if the holographic principle does not apply?}

One possibility is that the HP, for whatever reason, does not apply as a reliable proxy for the information content of a given physical system. It is worth reminding that the HP prescribes the maximum information content to scale with the surface, and not the volume, of a system: hence it generally already provides a very low information budget estimate, compared to all other proxies in which information scales with the volume, instead. In this sense, the estimates used in this paper (including in the low resolution simulation of Earth in Sec.\ref{low_res_earth}) already appears as conservatively low quantities.
If the HP is not valid, then a larger amount of bits can be encoded within a given surface with radius $R$ (in contradiction with our understanding of black hole physics). This will allow the usage of smaller computing devices, but at the same time it will require even more computing power, making the SH even more implausible.  
On the other hand, while the HP prescribes the maximum information which can be encoded with a portion of spacetime, the actual evolution of the enclosed system might be described with fewer bits of information. For example, in recent work \citet{va17info,va20complex} we used mathematical tools from Information Theory \citep[e.g.][]{adami,prokopenko2009information} to show that, on macroscopic scales, the statistical evolution of the cosmic web within the observable Universe can be encoded by using (only) $\sim 4 \cdot 3 \cdot 10^{16} \rm bits$ of information. This is of course incredibly less than the $I_U \sim 3.5 \cdot 10^{124} \rm bits$ estimate quoted in Sec.\ref{full_res_universe} and following from the HP. However, the latter includes all possible evolution on all scales, down to the $l_p$ fundamental length, in which a plethora of multi-scale phenomena obviously happens. While the concept of "emergence" and efficiency of prediction for complex multi-scale phenomena is a powerful and useful tool to compress the information needed to describe physical patterns forming at specific scales \citep[e.g.][]{shalizi2001causal}, a full error-less simulation of a multi-scale system seems to require a much larger amount of information. 
In essence, we shall conclude that while it is well conceivable that the actual information needed to fully capture the evolution of multi-scale phenomena emerging on different scales can be further reduced, it is implausible that  the information budget quoted in this work can be reduced by several orders of magnitude. 
\bigskip

\subsection{Plot twist: a simulated Universe simulates how the real universe might be} 
\label{simulations}
Our results suggest that no technological advancement will make the SH possible in any universe that works like ours. 

However, the limitations outlined above might be circumvented if the values of some of the fundamental constants involved in our formalims are radically different than the canonic values they have in this Universe.  Saying anything remotely consistent about the different physics operating in any other universe is an impossible task. And let alone, to guess which combinations of constants would still allow the development of any form of intelligent life.
Nevertheless, just for the sake of the argument, we make the very bold assumption that in each of explored variations some sort of intelligent form of life can form, and that it will be interested in computing and simulations. Under this assumption, we can then  explore numerical changes to the values of fundamental constants involved in the previous modelling, to see whether combinations exist, which can make at least a low resolution simulation of our Planet doable with a limited use of time and energy. 

So, in a final plot twist whose irony should not be missed, now this Universe (which might be a simulation) attempts to Monte-Carlo simulate how the  "real universe" out there could be, in order for the simulation of this Universe to be possible.
We assume that all known physical laws involved in our formalims are valid in all universes, but we allow each of the key fundamental constant to randomly vary across realisations.  For the sake of the exercise, we shall fix the total amount of information required for the low resolution simulation of our Planet discussed in Sec.\ref{low_res_earth} ($I_{\oplus,low}=1.65 \cdot 10^{51} \rm bits$) and consider the Hawking temperature relative to the black hole which is needed in each universe to perform the simulation. 

\begin{figure*}
    \centering
    \includegraphics[width=0.485\textwidth]{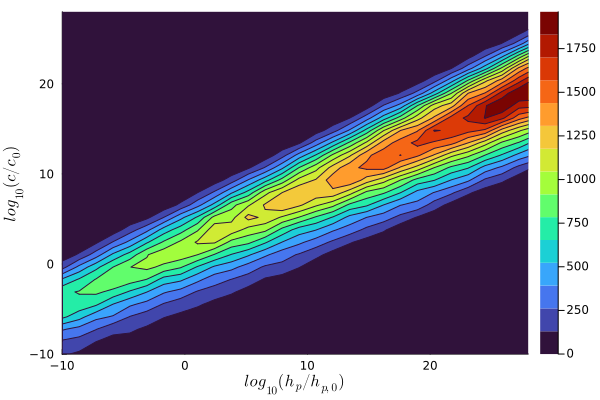}
    \includegraphics[width=0.485\textwidth]{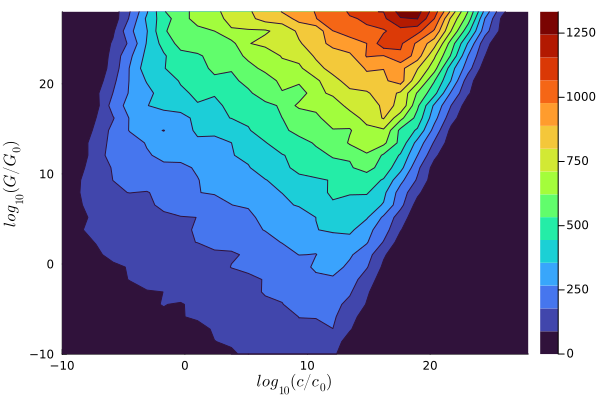}
    \includegraphics[width=0.485\textwidth]{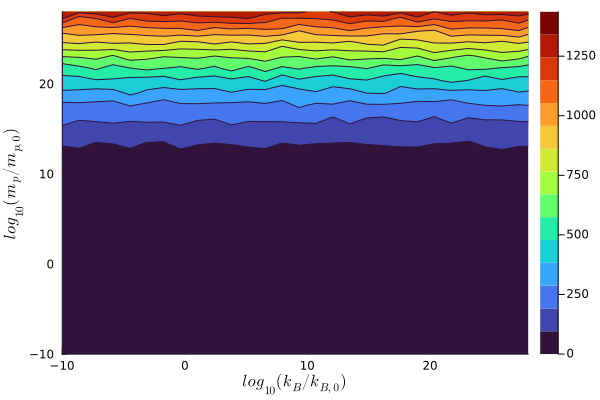} 
      \includegraphics[width=0.485\textwidth]{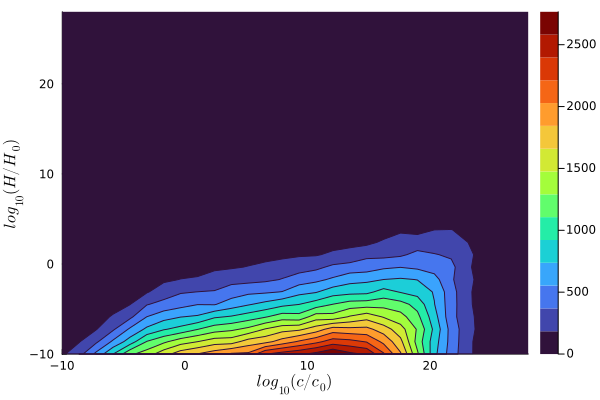}
          
    \caption{Examples of the distribution of allowed values of the fundamental constants (rescaled to their value in this Universe) necessary to perform a low-resolution simulation of Earth in other universes, as 
    predicted with Monte-Carlo simulations.
     Each axis gives the allowed value for each constant, normalised to the value it has in this Universe.}  \label{fig:sim}
\end{figure*}

The simulation consists in a Monte-Carlo exploration of the 6-dimensional parameter space of the fundamental constants which entered our previous derivation: $G$, $m_p$, $k_B$, $c$, $\hbar$, to which we now add $H$, i.e. the "Hubble-Lemaitre constant"  {\footnote{Very likely to be named in a different way in any other universe.}}  to compute the cosmic time. From $10^6$ randomly drawn universes, we select all realisations in which the low-resolution simulation of our Planet is "possible". Each constant is let free to randomly vary across $40$ orders of magnitudes around its reference value.
Our simplistic definition for possibility for the simulation here stems from two conditions: 

\begin{itemize}
\item the simulation can be run at least in real time, or faster than real time, in the universe where the simulation is produced. This means that a time interval of $1 \rm s$ in this Universe is simulated in the equivalent of  $1 \rm s$ in the other universe. However, any other universe can have different evolutionary timescales than this one, depending on their $H$. By considering that $1~ \rm s \approx 3 \cdot 10^{-18} t_U$, where $t_U$ is the age of this Universe, we require that one second of this Universe can be simulated in $\Delta t=3 \cdot 10^{-18} t_U$ of computing time in any other universe. To easily compute $t_U$, we assume for simplicity Einstein-de Sitter cosmology in every other universe, hence $t_U=2/(3 ~H)$, here $H$ is the Hubble-Lemaitre constant of other universes. 

\item The used power to produce the simulation is "reasonable". How can we guess what "reasonable" would be in any other universe? We cannot, of course, but for the sake of the exercise we use  $1 \rm ~GW$ of power as goal: this is about the power provided by a modern nuclear reactor, and it is $\sim 10^2-10^3$ higher than the typical power consumed by the best High Performance Computing centre to-date.  Thus it represents some extreme power budget available to  numerical astrophysicists of the remote future. 
As in the previous case, we must scale this power for the properties of other universes, 
and to do that we consider that $1 \rm ~GW$ equates to $\sim 6.6 \cdot 10^{18}$ protons ($\sim 10^{-33}$ of the mass of Earth) converted into energy in a second (where for the second, we again use the normalisation of the previous step). Therefore, we assume that in any other universe there are planets (or something conceptually equivalent) and that a very tiny fraction of their mass can be used to support computing. 
\end{itemize} 

We give in Fig.\ref{fig:sim} the results of the Monte-Carlo simulation, in which we show samplings of the distribution of allowed combinations of constants, normalised to the value each of them has in this Universe.
Although the exploration of the full 6-dimensional parameters space is complex, a few things can be noticed already.
First, two well correlated parameters here are $c$ and $h$: for a given value of the age of the universe, the computing timescale scales as $\propto h/T_H$, so  if $c$ increases, the Hawking radiation temperature of the black hole increases too (lowering the computing timescale), while at the same time proportionally larger values of $h$ are allowed. An other couple of correlated constants is  ($c$,$G$), which stems from the condition on the power: from this condition it follows that $c^3/G \propto m_p$, hence an order of magnitude of increase in $c$ must be compensated by a three orders of magnitude increase in $G$, for a fixed value of $m_p$. Most of the other combinations of constants (e.g. $(m_p,k_B)$ and $(H,c)$ in Fig.\ref{fig:sim}) show instead lacks of obvious correlation.

In general, this simulation shows that combinations of parameters exist, to make the SH for a low-resolution version of planet Earth possible (and similarly, also for higher resolution version of the SH), although they require orders of magnitudes in difference compared to the physical constants of this Universe. It clearly is beyond the goal of this work to further elaborate if some of the many possible combinations make sense, based on known physics, if they can support life and whether that hypothetical forms of life will be interested in numerics and physics, as we are {\footnote{The full direct simulation of such cases, possibly down to the resolution at which conscious entities will emerge out of the simulation and start questioning whether their universe is real or simulated, is left as a trivial exercise to the reader.}}. 

\bigskip

\subsection{What if the universe performing the simulation is entirely different than our Universe?}

Guessing how conservation laws for energy and information applies in a universe with entirely different laws, or whether they should even apply in the first place, appears impossible and this entirely prevents us from guessing whether the SH is possible in such case.
For example, hypothetical conscious creatures in the famous Pac-Man video game in the '80s will just be incapable of figuring out the constraints on the universe in which their reality is being simulated, even based on all the information they can gather around them. They would not guess the existence of gravity, for example, they would probably measure energy costs in "Power Pellets", and they would not conceive the existence of a third dimension, or of an expanding space time, and so on. 
Even if they could ever realise the level of graininess of their reality, and make the correct hypothesis of being living in a simulation, they would never guess how the real universe ("our" Universe, if it is real indeed) function in a physical sense. 
In this respect, our modelling shows that the SH can be reasonably well tested only with respect to universes which are at least playing according to the Physics play book - while everything else appears beyond the bounds of falsifiability and even theoretical speculation. 

%In our case, such graininess is yet to be discovered, and this contribution we showed how this consideration leads to deep physical constraints, which can only be eluded if the laws of the other universe simulating this one possess entirely different fundamental constants, or just are an entirely different set of laws. 

\section{Conclusions}
\label{conclusions}
We used standard physical laws to test whether this Universe, or some low resolution version of it, can be the product of a numerical simulation, as in the popular Simulation Hypothesis \citep[][]{Bostrom2003}.

One first key result is that the Simulation Hypothesis has plenty of physical constraints to fulfil, because any computation is bound to obey physics. We report that such constraints are so demanding, that all plausible approaches tested in this work (in order to reproduce at least a fraction of the reality we humans experience) require the access to impossibly large amounts of energy, or computing power. 
We are confident that this conclusively shows the impossibility of a "Matrix" scenario for the SH, in which our reality is a simulation produced by future descendants, machines or any other intelligent being in a Universe which exactly is at the one we (think we) live in - a scenario famously featured in the "The Matrix" movie in 1999, among many others.
We  showed that this hypothesis is just incompatible with all we know about physics, down to scales which have been already robustly explored by telescopes, high energy particle colliders or other direct experiments on Earth. 

What if our reality is instead the product of a simulation, in which physical laws (e.g. fundamental constants) different than in the reality are used?  A second result of this work is that, even in this alienating scenario, we can still robustly constrain the range of physical constants allowed in the reality simulating us.  In this sense, the strong physical link between computing and  energy also offers a fascinating way to connect hypothetical different levels of reality, each one playing according to the same physical rules. In our extremely simplistic Monte-Carlo scan of models with different fundamental constants, a plethora of combinations seems to exist, although we do not dare to guess which ones could be compatible with stable universes, the formation of planets and the further emergence of intelligent life. The important point is that each of such solutions implies universes entirely different from this one. 
Finally, the question whether universes with entirely different sets of physical laws or dimensionalities could produce our Universe as a simulation, seems to be be entirely outside of what is scientifically testable, even in theory. 

At this point, we shall notice that a possible "simulation hypothesis", which does not pose obvious constraints on computing, might be the solipsistic scenario in which the simulation simulates "just" the single activity of the reader's brain (yes: {\it you}), while all the rest is a sophisticated and very detailed hallucination. In this sense, nothing is new from Renee Descartes' "evil  genius" or "Deus deceptor" - i.e. for some reason an entire universe is produced in a sort of simulation, just to constantly fool us - from its more modern version of "Boltzmann Brains" \citep[][]{2010PhRvD..82f3520D}. Conversely, a more contrived scenario in which the simulation simulates only the brain activity of single individuals, appears to quickly run into the limitations of the SH exposed in this work: a shared and consistent experience of reality requires a consistent simulated model of the world, which quickly escalates into a too demanding model of planet Earth down to very small scales as soon as physical experiments are involved \footnote{A possibility for such model to work, may be that all scientists actually are similar to "non playable characters" in video games, i.e. roughly sketched and unconscious parts of the simulation, playing a pre-scripted role (in this case, reporting about fake measurements).}.

At this fascinating cross road between physics, computing and philosophy, it is interesting to notice that the last egotistic version of the SH appears particularly hard to test or debunk with physics, as the latter indeed appears to be entirely relying on the concept of a reality external to the observing subject - be it a real or a simulated one.  However, the possibility of a quantitative exercise like the one attempted here also shows that the power of physics is otherwise immense, as even the most outlandish and extreme proposal about our reality must fall within the range of what physics can investigate, test and debunk.  

Luckily, even in the way most probable scenario of all (the Universe is not a simulation), the amount of mysteries for physics to investigate still appears so immense, that even dropping this fascinating one cannot make science any less interesting.

\section{Additional Requirements}

\section*{Conflict of Interest Statement}

The authors declare that the research was conducted in the absence of any commercial or financial relationships that could be construed as a potential conflict of interest.

%\section*{%Author Contributions}
%F.V. is responsible for the writing of the manuscript and 

%\section*{Funding}
%The PI of this work has been supported by Fondazione Cariplo and Fondazione CDP, thorugh grant n° Rif: 2022-2088 CUP J33C22004310003 for "BREAKTHRU" project.
\section*{Acknowledgments}
I have used the Astrophysics Data Service (\href{http://adsabs.harvard.edu/abstract_service.html}{\tt ADS}) and \href{https://arxiv.org}{\tt arXiv} preprint repository extensively during this project and for writing this paper, the precious online  \href{http://www.astro.ucla.edu/~wright/CosmoCalc.html}{\tt Cosmological Calculator} by E. Wright, and the \href{https://julialang.org/}{\tt Julia} language for the simulations of Sec.\ref{simulations}. 
This peculiar investigation was prompted by a public conference given for the 40th anniversary of the foundation of \href{https://www.astrofili-vittorioveneto.it/index.php/2024/01/i-nostri-primi-40-anni/}{Associazione Astrofili Vittorio Veneto}, where the author took his first steps into astronomy.    
I wish to thank Maksym Tsizh, Chiara Capirini for their useful comments on the draft of the paper, and Maurizio Bonafede for useful suggestions on the geophysical analysis of the core-mantle boundary used in Sec.\ref{low_res_earth}.

\bibliographystyle{frontiersinHLTH&FPHY} % for Health, Physics and Mathematics articles
\bibliography{info,references}

\end{document}